# Poincaré's Dynamics of the Electron – A Theory of Relativity?

Galina Weinstein[*]

Before 1905, Poincaré stressed the importance of the method of clocks and their synchronization, but unlike Einstein, magnet and conductor (asymmetries in Lorentz's theory regarding the explanation of Faraday's induction) or chasing a light beam and overtaking it, were not a matter of great concern for him. In 1905 Poincaré elaborated Lorentz's electron theory from 1904 in two papers entitled "Sur la dynamique de l'electron". In May 1905 he sent three letters to Lorentz at the same time that Albert Einstein wrote his famous May 1905 letter to Conrad Habicht: "I can promise you in return four works, […] The fourth paper is only a rough draft at this point, and is an electrodynamics of moving bodies". In the May 1905 letters to Lorentz Poincaré presented the basic equations of his 1905 Dynamics of the Electron. Hence, in May 1905, Poincaré and Einstein both had drafts of papers pertaining to the principle of relativity. Poincaré's draft led to a space-time mathematical theory of groups at the basis of which stood the postulate of relativity, and Einstein's draft led to a kinematical theory of relativity. Poincaré did not renounce the ether. He wrote a new law of addition of velocities, but he did not abandon the tacit assumptions made about the nature of time, simultaneity, and space measurements implicit in Newtonian kinematics. Although he questioned absolute time and absolute simultaneity, he did not make new kinematical tacit assumptions about space and time. He also did not require reciprocity of the appearances, and therefore did not discover relativity of simultaneity: these are the main hallmarks of Einstein's special theory of relativity. Nevertheless, as shown by other writers, Poincaré's theory had influenced later scientists especially Hermann Minkowski.

## 1. 1905 – The Dynamics of the Electron

In 1905 Poincaré elaborated Lorentz's electron theory from 1904 in two papers entitled "Sur la dynamique de l'electron",[1] the first of which was a report and an outline of the latter. The report was presented on June 5, 1905 to the French Academy of Sciences and published in the *Comptes rendus hebdomadaires des seances de l'Academie des sciences. Paris*, 4 pages long, the usual length of reports published in the *Comptes rendus*. On July 23, 1905 Poincaré submitted the second paper – published only in 1906 – to an obscure Italian journal by the name, *Rendiconti del Circolo matematico di Palerno*.[2]

During May 1905 Poincaré sent three undated letters to Lorentz, in which he presented to the latter the essential elements of his theory. At the same time, however, it is a curious coincidence that Albert Einstein wrote his famous May 1905 undated letter to Conrad Habicht, "I can promise you in return four works, […] The fourth paper is only a rough draft at this point, and is an electrodynamics of moving bodies which employs a modification of the theory of space and time; the purely kinematical part of this paper will surely interest you".[3]

---

[*]Written while at the Center for Einstein Studies, Boston University

In May 1905, both Poincaré and Einstein possessed drafts of papers pertaining to the principle of relativity. However, Einstein was developing a new kinematics leading to a theory of relativity, and Poincaré perfected Lorentz's theory using his mathematical theory of groups to a stage that one could not *at all* disclose absolute motions. Yves Pierseaux described the difference between the two theories,[4]

"The two Famous papers, that of Einstein ("Elektrodynamik bewegter Körper") and that of Poincaré ("La dynamique de l'elrctron") contain not only two approaches of SR but two different *theories* of SR".[5] And Pierseaux explains what he means by two different theories, Einstein's principles of kinematics and Poincaré's theory of groups.

Let us examine Poincaré's three letters to Lorentz before discussing Poincaré's paper. The letters reveal the manner and succession in which he originally presented the equations; then we can compare this presentation to that appearing in Poincaré's paper.

## 2. The May 1905 Letters to Lorentz

*In the first letter* that was sent sometime during May 1905, Poincaré began by telling Lorentz,[6]

"For some time I have studied in greater detail your [1904] memoir electromagnetic phenomena in a system moving with any velocity smaller than that of light, the importance of which is extreme, and I have already mentioned the main results at the conference in St. Louis".

What were the main results? In the lecture he gave on September 24 1904 at a congress for arts and science in Saint Louis, "L'État Actual et l'Avenir de la Physique Mathématique" (The Present State and Future of Mathematical Physics – later published in *Bulletin des sciences mathématiques*), Poincaré explained his physical interpretation of Lorentz's local time by synchronization of clocks by light signals:[7]

Poincaré starts with two observers who exchange signals, but they know that the transmission of light is not instantaneous, and they cross the signals.

When the station B perceives the signal from the station A, its clock should not mark the same hour as that of the station A at the moment of sending the signal. This hour is augmented by a constant representing the duration of the transmission.

Suppose that the station A sends its signal when its clock marks the hour zero, and that the station B perceives it when its clock marks the hour t. The clocks are synchronized if the slowness is equal to t represents the duration of the transmission, and to verify this, the station B sends in its turn a signal when its clock marks zero. The station A then should perceive it when its clock marks t. The watches are then synchronized.

And, in fact, they mark the same hour at the same physical instant, but on one condition, which is that the two stations are fixed.

If the two stations are moving, then the duration of transmission will not be the same in both directions, since the station A, for example, moves forward to meet the optical perturbation emanating from B, while the station B flies away before the perturbation emanating from A.

The watches synchronized in that manner do not mark, therefore, the true time, they mark the *local time*, so that one of them goes slow on the other. It matters little, since we have no means of perceiving it. All the phenomena which happen at A, for example, will be late, but all will be equally so, and the observer who ascertains them will not perceive it since his watch is slow.

Poincaré could not require reciprocity of the appearances, i.e., the velocity of the frame "at rest" relative to the moving system is equal and opposite to that of the moving system relative to the system at rest. For Poincaré, in the moving system an observer is measuring "apparent" lengths and "apparent" time units, and in the ether frame he is measuring "real" lengths and "real" time units.

As opposed to Einstein, before 1905 Poincaré stressed the importance of the method of clocks and their synchronization by light signals.[8] Poincaré did not consider Faraday's induction (asymmetries in Lorentz's theory regarding the explanation of Faraday's induction) or catching up with a light beam and overtaking it, as a matter of great concern for him. For Einstein the latter were crucial while synchronization of clocks by light signals did not play an important role in his process of discovery.[9]

After writing in the letter that he had been studying in great detail Lorentz's 1904 work, Poincaré wrote to Lorentz about the following discovery,[10]

"I agree with you on all essential points; however, there are some differences.

Hence on page 813, it is set:[11]

$$\frac{1}{kl^3}\rho = \rho'; \; k^2 u_x = u'_x, k^2 u_y = u'_y$$

It seems to me that one should set,

$$\frac{1}{kl^3}(\rho(1+\varepsilon v_x) = \rho'. \quad \frac{1}{kl^3}\rho(v_x+\varepsilon) = \rho' u'_x$$

where, $\varepsilon = -\frac{\omega}{c}$ or $\varepsilon = -\omega$ if we choose the units in such a way that c = 1.

The change seems to me necessary if one wants the apparent charge of the electron to be conserved".

Subsequently Poincaré modified two more equations appearing on page 813 of Lorentz's paper.

Lorentz answered Poincaré, but his letter was not preserved.[12]

*The second letter* of Poincaré to Lorentz was sent some time during May 1905. Poincaré wrote, "Thank you for your kind letter".[13]

Poincaré now reported to Lorentz, "Ever since I have written my idea there are a few points that have changed. I have found like you by another route that $l = 1$".[14] What was this route? From Poincaré's paper[15] we know that he used group theory to demonstrate that $l = 1$. This was a great discovery which paved the way to Poincaré's Lorentz group.

Poincaré then went straight to his major discovery: he sent Lorentz the correct coordinate and time transformations (Lorentz transformations). Poincaré corrected Lorentz's 1904 transformations. Lorentz wrote in his 1904 paper on page 812,[16]

$$k^2 = \frac{c^2}{c^2 - \omega^2},$$

$$x' = klx, \quad y' = ly, \quad z' = lz, \quad t' = \frac{l}{k}t - kl\frac{\omega}{c^2}x.$$

($\omega$ is the velocity of translation).

Poincaré wrote Lorentz,[17]

"Let $-\varepsilon$ be the velocity of translation, so that of light is taken to be unity.

$$k = (1 - \varepsilon^2)^{-\frac{1}{2}}$$

We get the transformation,

$$x' = kl(x + \varepsilon t), \quad t' = kl(t + \varepsilon x),$$
$$y' = ly, \quad z' = lz,$$

These transformations form a group".

This is the first time that Poincaré spoke about the famous *Lorentz group*. And subsequently Poincaré demonstrated to Lorentz that this was indeed the case,

"Consider two transformations, the components of which correspond to

$k, l, \varepsilon$

and

$k', l', \varepsilon'$

they result in corresponding to

k″, l″, ε″

where:

$$k'' = \left(1 - \varepsilon''^2\right)^{-\frac{1}{2}}, l'' = ll',$$

$$\varepsilon'' = \frac{\varepsilon + \varepsilon'}{1 + \varepsilon\varepsilon'}$$

If we now put

$$l = (1 - \varepsilon^2)^m, l' = (1 - \varepsilon'^2)^m$$

We will get:

$$l'' = (1 - \varepsilon''^2)^m$$

Because for m = 0."

Poincaré sent Lorentz another letter in May, 1905, *the third letter*. This time he did not send him any equations and mathematical derivations. He told Lorentz that he was continuing his research. The upshot of the letter was, "My results fully confirm yours in the sense that the compensation is perfect (which prevents the experimental determination of absolute motion) and can only be complete with the hypothesis $l = 1$".[18] Using group theory Poincaré managed to demonstrate that the postulate of relativity was *fully valid* in Lorentz theory.

Less than a month later, Poincaré reported to the French Academy of Sciences about his discovery, and he published the four pages June 4 1905 note "On the Dynamics of the Electron" in the *Comptes rendus*.

**3. Introducing the Problems**

In the introduction of the paper, especially the extended version, Poincaré presented the problems that occupied him. The introduction explained the motif for taking the study: why did Poincaré write a long paper centered on one postulate, the postulate of relativity (embodied in the Lorentz transformations)?

Poincaré began his paper by explaining this,[19]

"It seems at first that the aberration of light and related optical and electrical phenomena will provide us with a means of determining the absolute motion of the Earth, or rather its motion with respect to the ether, as opposed to its motion with respect to other celestial bodies. Fresnel pursued this idea, but soon recognized that the Earth's motion does not alter the laws of refraction and reflection. Analogous experiments, like that of the water-filled telescope, and all those considering terms no higher than first order relative to the aberration, yielded only negative results; the explanation was soon discovered. But Michelson, who conceived an experiment sensitive to terms depending on the square of the aberration, failed in turn. It appears that this impossibility to detect the absolute motion of the Earth by experiment is a general law of nature; we are naturally led to admit this law, which we will call the *Postulate of Relativity* and admit without restriction.[20] Whether or not this postulate, which up to now agrees with experiment, may later be corroborated or disproved by experiments of greater precision, it is interesting in any case to ascertain its consequences".

The results of Michelson's ether drift experiments pointed towards the postulate of relativity, and very soon,[21]

"An explanation was proposed by Lorentz and FitzGerald, who introduced the hypothesis of a contraction of all bodies in the direction of the Earth's motion and proportional to the square of the aberration. This contraction, which we will call the *Lorentzian contraction*, would explain Michelson's experiment and all others performed up to now. The hypothesis would become insufficient, however, if we were to admit the postulate of relativity in full generality".

Poincaré said that in order to achieve an agreement with the principle of relativity, Lorentz complemented the contraction hypothesis by demonstrating,[22]

"If we are able to impress a translation upon an entire system without modifying any observable phenomena, it is because the equations of an electromagnetic medium are unaltered by certain transformations, which we will call *Lorentz transformations*. Two systems, one of which is at rest, the other in translation, become thereby exact images of each other".

Already in 1900 Poincaré objected to the invention of a new hypothesis every time a negative result was received as an outcome of a newly performed experiment. As far as he was concerned, it was not mere chance that all the experiments had so far led to a negative result; this must have signified a rule of nature, imposed upon them, according to which, in principle, no such experiment could ever give a positive result,[23]

"Experiments were performed that should have detected the terms of the first order; the results of which were negative; could it be a mere chance? [...] Then more precise experiments were performed, they were also negative; could this be too a result of chance [?];"

Poincaré searched for a perfect compensation that would prevent us from ever detecting motion with respect to the ether, which he considered a convenient hypothesis.

During the academic year 1888, Poincaré taught his students in the second semester of 1887-1888 in the faculty of sciences in Paris the mathematical theory of light, *cours de physique mathématique Théorie mathématique de la lumière*. Poincaré later organized his lectures (probably transcript by his students), and in the introduction from December 2$^{nd}$, 1888, he wrote: [24]

"It matters to us little whether the ether really exists; it is the matter of metaphysicians; what is essential for us is that everything happens as if it existed and that this hypothesis is convenient for the explanation of phenomena. After all, have we any other reason for believing in the existence of material objects? That too is only a convenient hypothesis; only it will never cease to be so, while a day will come no doubt in which the ether will be rejected as useless".

Two years later, on August 6, 1900, Poincaré participated in the international congress of physics in Paris. There were many participants from the entire world, and there was an exhibition. Poincaré gave the keynote lecture "Sur les rapports de la physique expérimentale et de la physique mathématique" (Relations between Experimental and Mathematical Physics) in the mathematics session of the congress.

In the printed version of the talk he asked, "And our ether, does it really exist?", and then added: [25] "We know whence comes our belief in the ether. […] Fizeau's experiment goes further. By the interference of rays that traverse through the air or the water in movement, it seems we watch two different media penetrating each other and yet moving with respect to each other. One believes to touch the ether with the finger". Did Poincaré accept in 1900 Fresnel's interpretation for the dragging coefficient? It is important to stress that Armand-Hippolyte Fizeau only confirmed Fresnel's formula, *not the explanation in terms of partial ether drag given by Fresnel to this formula.*[26]

The French editor, Camille Flammarion, asked Poincaré to collect his papers into a general philosophical volume accessible to the general reader. In 1902 Poincaré combined in his book *La Science et l'hypothèse* (Science and Hypothesis) his 1888 and 1900 scientific works, including the comments, which were comprehensible for the general reader, as expressed in two completely different contexts.[27] In his book Poincaré formulated the principle of relative motion, spoke of ether drift experiments, and then these two paragraphs of 1888 and 1900 formed a coherent line of thought according to which the ether might be useless. *However, Poincaré never gave up the ether*, he objected to mechanistic models underlying the ether because,[28]

"*If, then a phenomenon involves a complete mechanical explanation, it will involve an infinite of others that will equally well report all the particularities revealed by experience*".

Poincaré ended the 1905 Dynamics of the Electron introduction by saying,[29]

"If we were to admit the postulate of relativity, we would find the same number in the law of gravitation and the laws of electromagnetism – the speed of light – and we would find it again in all other forces of any origin whatsoever. This state of affairs may be explained in one of two ways: either everything in the universe would be of electromagnetic origin, or this aspect – shared, as it were, by all physical phenomena – would be a mere epiphenomenon, something due to our methods of measurement".

In his 1905 Dynamics of the Electron Poincaré *did not formulate the constancy of the speed of light as a postulate*. He very likely objected to such a postulate, and he only accepted the relativity principle as a postulate.

Poincaré then alluded to the method of clocks and their synchronization by light signals, [30]

"How do we go about measuring? The first response will be: we transport objects considered to be invariable solids, one on top of the other. But that is no longer true in the current theory if we admit the Lorentzian contraction. In this theory, two lengths are equal, by definition, if they are traversed by light in equal times".

Like Einstein Poincaré adopted a definition of distant simultaneity. However, unlike Einstein, Poincaré did not discover the relativity of simultaneity. In 1902, Poincaré wrote a letter to the Royal Academy of Sciences in Stockholm recommending the candidacy of Lorentz for a Nobel Prize in Physics. In trying to persuade the Nobel committee about Lorentz's achievements, Poincaré wrote the following,[31]

"Why for example all the experiments devoted to demonstrating the Earth's motion gave negative results? Evidently, there was one general reason behind this; this reason was discovered by Mr. Lorentz and he put it in a striking form with his ingenious invention of 'reduced time'. Two phenomena taking place in two different places can appear simultaneous even though they are not: everything happens as if the clock in one of these places retards with respect to that of the other, and as if no conceivable experiment could show evidence of this discordance. Now, according to Mr. Lorentz, the effect of the Earth's motion would be only to give rise to a similar discordance that no experiment could reveal".

John Stachel explained,[32]

"Poincaré had interpreted the local time as that given by clocks at rest in a frame moving through the ether when synchronized as if – contrary to the basic assumptions of Newtonian kinematics – the speed of light were the same in all inertial frames. Einstein dropped the ether and the 'as if': one simply synchronized clocks by the Poincaré convention in each inertial frame and accepted that the speed of light really is the same in all inertial frames when measured with clocks so synchronized".

Although Poincaré did not discover relativity of simultaneity *he was the first to question absolute simultaneity and absolute time*. In 1900 in a talk at the Paris Philosophy congress, "On the Principles of Mechanics", Poincaré wrote,[33]

"1. There is no absolute space and we only perceive relative movements; however one expresses most often mechanical facts as if there was an absolute space to which they could be referred.

2. There is no absolute time; saying that two durations are equal is an assertion that has no meaning to it and can only be acquired one by convention.

3. Not only do we have no direct intuition of the equality of two durations, but we do not even have it of the simultaneity of two events which are produced in two different scenes; this is what I explained in an article entitled the *Measurement of Time*[1]". [34]

In a footnote Poincaré gave exact reference to this paper. The above 1900 passage reappeared in Poincaré's 1902 book *Science and Hypothesis*.[35]

*Poincaré did not renounce the ether. He wrote a new law of addition of velocities, but he did not abandon the tacit assumptions made about the nature of time, simultaneity, and space measurements implicit in Newtonian kinematics. Although he questioned absolute time and absolute simultaneity, he did not make new kinematical tacit assumptions about space and time. He also did not require reciprocity of the appearances, and therefore did not discover relativity of simultaneity: these are the main hallmarks of Einstein's special theory of relativity*

## 4. The Lorentz Transformations

Section §1 and section §4 of Poincaré's paper, "On the Dynamics of the Electron", contain material directly pertaining to the principle of relativity, and section §9 discusses gravitation. Sections §6 to §8 discuss the configuration of the electron and present the theory of the Poincaré pressure. The Dynamics of the Electron is a mathematical physics theory, rather than a theoretical physics one.[36]

In section §1 Poincaré presented the main results pertaining to the postulate of relativity. He started with the Maxwell-Lorentz fundamental equations – which he designated by (1) and the equation for the Lorentz force, designated by (2). He then showed that "These equations admit a remarkable transformation discovered by Lorentz" (in 1904). These were the complete Lorentz transformations that Poincaré had written in his second May 1905 letter to Lorentz,[37]

$$(3) \quad x' = kl(x + \varepsilon t), \quad t' = kl(t + \varepsilon x), \quad y' = ly, \quad z' = lz,$$

where, $l$ and $\varepsilon$ are two arbitrary constants, and,

$$k = \frac{1}{\sqrt{1-\varepsilon^2}}.$$

Poincaré *postulated* these transformations and did not derive them. He explained that the Lorentz transformation "owes its interest to the fact that it explains why no experiment can inform us of the absolute motion of the universe".[38]

Poincaré then considered a sphere which was carried along with the electron in uniform translation,[39]

$$(x - \xi t)^2 + (y - \eta t)^2 + (z - \zeta t)^2 = r^2,$$

where, ξ, η, ζ are the velocity components of the electron. The volume of the sphere is, $\frac{4}{3}\pi r^3$.

The Lorentz transformation (3) changes the sphere into an ellipsoid. This ellipsoid is in uniform motion. Its volume for t' = 0 is, [40]

$$\frac{4}{3}\pi r^3 \frac{l^3}{k(1+\xi\varepsilon)}.$$

The charge of the electron is invariant under the Lorentz transformations (3). Poincaré designated the new charge density by ρ' and wrote, [41]

$$(4)\ \rho' = \frac{k}{l^3}\rho(1+\varepsilon\xi).$$

He then wrote the new velocity components ξ', η', ζ' and obtained the "Règle d'addition des vitesses", the addition law for velocities, [42]

$$\xi' = \frac{dx'}{dt'} = \frac{d(x+\varepsilon t)}{d(t+\varepsilon x)} = \frac{\xi+\varepsilon}{1+\varepsilon\xi}.$$

$$\eta' = \frac{dy'}{dt'} = \frac{dy}{kd(t+\varepsilon x)} = \frac{\eta}{k(1+\varepsilon\xi)}.$$

$$\zeta' = \frac{dz'}{dt'} = \frac{\zeta}{k(1+\varepsilon x)}.$$

Then he wrote the corrected charge density transformations, [43]

$$(4')\ \rho'\xi' = \frac{k}{l^3}\rho(\xi+\varepsilon),\qquad \rho'\eta' = \frac{k}{l^3}\rho\eta,\qquad \rho'\zeta' = \frac{k}{l^3}\rho\zeta.$$

In the first letter to Lorentz, Poincaré wrote equations (4) and (4'), and only in the second letter he wrote equations (3). Poincaré did not write in his letters to Lorentz the addition law for velocities. Writing the "Règle d'addition des vitesses" in the 1905 paper reveals how Poincaré might have obtained in the first place equations (4) and (4'). According to the presentation in the 1905 paper, it is reasonable to assume that Poincaré had discovered the Lorentz transformations shortly *before* he found equations (4) and (4'). The interesting question is why did Poincaré send Lorentz equations (4) and (4') first, before sending him equations (3)? The reasonable answer is that Poincaré was not yet sure whether his Lorentz transformations (3) were actually correct, and he waited first for Lorentz's approval for his equations (4) and (4').

In his 1905 paper Poincaré proved that equations (4) and (4') satisfied the continuity condition.

Subsequently, using (3) Poincaré derived, by differentiation and using "formulae", which "are notably different from those of Lorentz" (transformation formula for the vector and scalar potentials) the transformation for the electric and magnetic fields, [44]

(9) $f' = \frac{1}{l^2}f, \quad g' = \frac{k}{l^2}(g + \varepsilon\gamma), \quad h = \frac{k}{l^2}(h - \varepsilon\beta),$

$\alpha' = \frac{1}{l^2}\alpha, \quad \beta' = \frac{k}{l^2}(\beta - \varepsilon h), \quad \gamma = \frac{k}{l^2}(\gamma + \varepsilon g),$

where, f, g, h are the electrical displacement, and α, β, γ, are the magnetic force. Poincaré then transformed Lorentz's force using equations (3).

With this ended section §1. In sections §2 and §3 Poincaré occupied himself with the principle of least action and the Lorentz transformation.

In section §4 Poincaré demonstrated that the Lorentz transformation forms a group, "We are thus led to consider a continuous group, which we call *the Lorentz group*".[45] Poincaré elaborated the demonstration that he had sent in the second letter to Lorentz in May 1905.[46]

Consider the Lorentz transformations

(3) $x' = kl(x + \varepsilon t), \quad t' = kl(t + \varepsilon x), \quad y' = ly, \quad z' = lz.$

And,

$x'' = k'l'(x' + \varepsilon't'), \quad t'' = k'l'(t' + \varepsilon'x'), \quad y'' = l'y', \quad z'' = l'z',$

with,

$k^{-2} = 1 - \varepsilon^2, \quad k'^{-2} = 1 - \varepsilon'^2,$

It follows that,

$x'' = k''l''(x + \varepsilon''t), \quad t'' = k''l''(t + \varepsilon''x), \quad y'' = l''y, \quad z'' = l''z,$

with,

$\varepsilon'' = \frac{\varepsilon + \varepsilon'}{1 + \varepsilon\varepsilon'}, \quad l'' = ll', \quad k'' = kk'(1 + \varepsilon\varepsilon') = \frac{1}{\sqrt{1 - \varepsilon''^2}},"$

Poincaré described the mathematical properties of the Lorentz group and then said, "Any transformation of this group can always be decomposed into a transformation having the form,[47]

$x' = lx, \quad y' = ly, \quad z' = lz, t' = lt$

and a linear transformation which leaves unaltered the quadratic form

$$x^2 + y^2 + z^2 - t^2.\text{"}$$

Poincaré did not associate this quadratic form with propagation of light in order to define a null interval like Einstein or a metric like Minkowski. Physics meaning of the quadratic form is not discussed by Poincaré.[48]

Recall that Poincaré reported to Lorentz in the second letter, "Ever since I have written my idea there are a few points that have changed. I have found like you by another route that $l = 1$".[49] In the 1905 paper Poincaré demonstrated this,[50]

(1) $\quad x' = kl(x + \varepsilon t), \quad y' = ly, \quad z' = lz, \quad t' = kl(t + \varepsilon x)$,

and preceding and following by an appropriate rotation. Poincaré says that, for our purpose we consider only part of the transformation in this group; and we consider $l$ as function of $\varepsilon$, in such a manner that this sub-group P is itself also a group.

Poincaré rotates the system through $180°$ about the y axis, the resulting transformation of which belongs to the sub-group P. The operation is tantamount to changing the signs of x, x', z, and z'. Hence we obtain,

(2) $\quad x' = kl(x - \varepsilon t), \quad y' = ly, \quad z' = lz, \quad t' = kl(t - \varepsilon x)$,

Thus $l$ is unchanged when $\varepsilon$ is changed by $-\varepsilon$.

Now since P is a group, then the substitution inverse of (1) is also a group:

(3) $\quad x' = \dfrac{k}{l}(x - \varepsilon t), \quad y' = \dfrac{y}{l}, \quad z' = \dfrac{z}{l}, \quad t' = \dfrac{k}{l}(t - \varepsilon x)$,

which belongs to P. And this last group is identical to (2), and thus,

$$l = \dfrac{1}{l}$$

and consequently we obtain, $l = 1$.

In section §9 when discussing his theory of gravitation, Poincaré extended his mathematical theory of groups from electrodynamics to gravitation.

*I wish to thank Prof. John Stachel from the Center for Einstein Studies in Boston University for sitting with me for many hours discussing special relativity and its history. Almost every day, John came with notes on my draft manuscript, directed me to books in his Einstein collection, and gave me copies of his papers on Einstein, which I read with great interest*

[1] Poincaré, Henri (1905b), "Sur la dynamique de l'électron", *Comptes-rendus des séances de l'Académie des sciences* 140, 5 Juin, 1905, pp. 1504-1508, Poincaré, Henri (1905c), "Sur la dynamique de l'électron (received on 23 July,1905), *Rendiconti del Circolo Matematico di Palermo* 21, 1906, pp. 129-175 (1-47).

[2] Poincaré, 1905c, p. 1. It was not the first time that Poincaré published papers in this journal. For instance: Poincaré, Henri, "Sur les équations de la physique mathématique", *Rendiconti del circolo matematico di Palermo* 8, 1894, pp. 57-156.

[3] Einstein to Habicht, 18 or 25 May, 1905, *The Collected Papers of Albert Einstein. Vol. 5: The Swiss Years: Correspondence, 1902–1914*, Klein, Martin J., Kox, A.J., and Schulmann, Robert (eds.), Princeton: Princeton University Press, 1993, Doc. 27.

[4] Pierseaux, Yves, "The 'fine structure' of Special Relativity and the Thomas precession", *Annals de la Fondation Louis de Broglie* 29, 2004, pp 57-116; p. 60.

[5] Pierseaux, 2004, p. 58.

[6] Poincaré to Lorentz, May, 1905, Poincaré Archives Nancy; Letter Number 38.3., Walter, Scott, Bolmont Étienne, and Coret, André, *La Correspondance entre Henri Poincaré et les physiciens, chimistes et ingénieurs*, 2000, Berlin: Birkhäuser, p. 255.

[7] Poincaré, Henri, "L'état actuel et l'avenir de la Physique mathématique", *Bulletin des sciences mathématiques*, 28, Décembre 1904, pp. 302-324; translated to English by G.B. Halsted: "The Principles of Mathematical Physics", *The Monist* 15, 1905, p. 1, p. 311. Three scientists from the French Academy of Sciences attended the Congress: Gaston Darboux, Émile Picard, and Poincaré.

Already in 1900, in a lecture held at the Lorentz Festschrift celebrations, "La théorie de Lorentz et le principe de reaction", Poincaré suggested a physical interpretation of Lorentz's local time by synchronization of clocks by light signals, Poincaré, Henri (1900b), "La théorie de Lorentz et le principe de réaction", *Archives néerlandaises des sciences exactes et naturelles. Recueil de travaux offerts par les auteurs à H.A.Lorentz* (The Hague: Nijhoff) Ser II, 1900, 5, pp. 252-278, p. 272: "For the compensation to be done, the phenomena should be related, not to the true time $t$, but to a certain *local time t'* defined in the following manner. I suppose that several observers placed at different points, synchronize their watches by means of light signals, and that they try to correct by these signals the times of transmission, but being ignorant of the translation motion in which they are moving, and believing as a consequence that the signals are transmitted equally fast in both directions, they confine themselves to crossing the observations, by sending one signal from A to B, then another from B to A. The local time t' is the time marked by the watches thus synchronized"

[8] Poincaré, Henri, "La mesure du temps, *Revue de métaphysique et de morale* 6, 1898, pp. 371-384, Poincaré, 1900b.

[9] My paper on Einstein's pathway to special relativity in this ArXiv; Norton, John, "Einstein's investigations of Galilean covariant Electrodynamics prior to 1905", *Archive for the History of Exact Sciences* 59, 2004, pp. 45-105; Norton, John, "Discovering the Relativity of Simultaneity: How Did Einstein Take 'The Step'," Trans. to Chinese, Wang Wei. In *Einstein in a Trans-cultural Perspective*. Eds. Yang Jiang, Liu Bing., 2008, Tsinghua University Press; Stachel, John, "'What Song the Syrens Sang': How Did Einstein Discover Special Relativity", Umberto Curi (ed.). Ferrara, Gambriele Corbino & Co., *L'Opera di Einstein*, 1989, pp. 21-37; English text of Spanish original in Stachel, John, *Einstein from 'B' to 'Z'*, 2002, Washington D.C.: Birkhauser, pp.157-170; Stachel, John, "Einstein on the Theory of Relativity", Headnote in *The Collected Papers of Albert Einstein. Vol. 2: The Swiss Years: Writings, 1900–1909*, Stachel, John, Cassidy, David C., and Schulmann, Robert (eds.), Princeton: Princeton University Press, 1989, pp. 253-274; reprinted in Stachel (2002), pp. 191-214; Stachel, John, "Einstein's Clocks, Poincare's Maps/Empires of Time", *Studies in History and Philosophy of Science* B 36 (1), 2005, pp. 202-210; pp. 1-15; Stachel, John, "A world Without Time: the Forgotten Legacy of Gödel and Einstein", *Notices of the American Mathematical Society* 54, 2007, pp. 861-868 (1-8); Stachel, John, "Albert Einstein", *The New Dictionary of Scientific Biography*, Vol 2, Gale 2008, pp. 363-373; Stachel, John, "Poincaré and the Origins of Special Relativity", PSA Meeting San Diego, November, 2011, pp. 15-17.

[10] Poincaré to Lorentz, May, 1905, Poincaré Archives Nancy; Letter Number 38.3., Walter, Bolmont, and Coret (ed), 2000, p. 255.

[11] Lorentz, Hendrik Antoon, "Electromagnetic Phenomena in a System Moving with any Velocity Smaller than that of Light", *Verslagen Konignklijke Akademie Van Wetenschapen (Amsterdam). Proceedings of the section of science,* 6, 1904, pp. 809-836; p. 813.

[12] Walter, Bolmont, and Coret (ed), 2000, p. 258, note 1.

[13] Poincaré to Lorentz, May, 1905, Poincaré Archives Nancy; Letter Number 38.4., Walter, Bolmont, and Coret (ed), 2000, p. 257.

[14] Poincaré to Lorentz, May, 1905, Poincaré Archives Nancy; Letter Number 38.4., Walter, Bolmont, and Coret (ed), 2000, p. 257.

[15] Poincaré, 1905c, p. 18.

[16] Lorentz, 1904, p. 312.

[17] Poincaré to Lorentz, May, 1905, Poincaré Archives Nancy; Letter Number 38.4., Walter, Scott, Bolmont Étienne, and Coret, André, *La Correspondance entre Henri Poincaré et les physiciens, chimistes et ingénieurs*, 2000, Berlin: Birkhäuser, p. 257.

[18] Poincaré to Lorentz, May, 1905, Poincaré Archives Nancy; Letter Number 38.5., Walter, Bolmont, and Coret (ed), 2000, p. 258.

[19] Poincaré, 1905c, p. 1.

[20] In August 1900, in the congress of philosophy, Poincaré formulated the Galilean principle of relative motion during his talk "On the Principles of Mechanics". This formulation was reproduced two years later in the chapters discussing Mechanics in *Science and Hypothesis*, "The motion of some system has to obey the same laws, whether with respect to fixed axes, or to mobile axes carried by a rectilinear and uniform motion. This is the principle of relative motion, which is imposed on us because of two reasons: first the most vulgar experiments have confirmed it, and also the contrary hypothesis is singularly repugnant to the mind". Poincaré, Henri (1900c), "Sur les principes de la mécanique (lecture before the international congress of philosophy in Paris, 1st, August 1900), *Bibliothèque du Congrèss international de philosophie* tenu à Paris du 1$^{er}$ au 5 aout 1900,Vol. III: *Logique et histoire des sciences*, 1901, Paris: Colin, pp. ,457-494; p. 477; Poincaré, Henri, *La science et l'hypothése*, 1902/1968, Paris: Flammarion (The second edition of 1968 is based on the one edited by Gustave Le Bon,1917); translated to English by W.J.G., *Science and Hypothesis* New York: The Walter Scott Publishing co, Feb 1905, New-York: Dover;1952; the French 1968 edition, p. 129. Four years later in the Saint Louis congress Poincaré spoke about the principle of relativity, Poincaré, 1904, p. 306. "The principle of relativity, according to which the laws of physical phenomena should be the same, whether for a fixed observer, or for an observer carried along in a uniform movement of translation; so that we do not have and could not have any means of discerning whether or not we are carried along in such a movement".

[21] Poincaré, 1905c, p. 1.

[22] Poincaré, 1905c, p. 2.

[23] Poincaré, Henri (1900a), "Sur les relations entre la physique expérimentale et la physique mathématique", *Revue génerale des sciences pures et appliqués* 21, 1900, pp. 1163-1175; p. 1172.

[24] Poincaré, Henri, *Lecons sur la théorie mathématique de la lumière, professées pendant le premier semestre 1887-1888*, Cours de physique mathématique, edited by Jules Blondin, Cours de la Faculté des sciences de Paris, 1889, Paris: Carré et Naud*semestre 1887-1888*, Cours de physique mathématique, edited by Jules Blondin,Cours de la Faculté des sciences de Paris, 1889, Paris: Carré et Naud, pp. I-II.

[25] Poincaré, 1900a, pp. 1171-1172.

[26] Fizeau, Armand-Hippolyte, "Sur les hypotheses relatives a l'ether lumineux, et sur une experience qui parait demontrer que le mouvement des corps change la vitesse avec laquelle la lumiere se propage dans leur interieur", *Comptes Rendus de l'Académie des Sciences* 33, 1851, pp. 349-355.

[27] Poincaré, 1902/1968, pp. 180, 215.

[28] Poincaré, Henri, *Electricité et optique, II. Les théories de Helmholtz et les expériences de Hertz. Lecons professées pendant le second semestre 1889-1890*, Cours de la Faculté des sciences de Paris, Cours de physique mathématique edited by Bernard Brunhes, 1891, Paris: Carré et Naud, p. xiv.

[29] Poincaré, 1905c, p. 2.

[30] Poincaré, 1905c, p. 2.

[31] Poincaré, Henri, "Nomination letter for the Nobel Prize in physics, Archives, Royal Academy of Science, Stockholm", a copy at the Archives of Henri Poincaré, Nancy, 1902.

"pourquoi par example, toutes les experiences tentées pour metre en evidence le mouvement de la Terre, ont-elles données des resultants negatives? Il était evident qu'il y avait à cela une raison générale; cette raison, M. Lorentz l'a découverte et il l'a mise sous une forme frappante par son ingénieuse invention du 'temps réduit'. Deux phénomènes qui se passent en deux lieux different peuvent paraître simultanés bien qu'ils ne le soient pas: tout se passe comme si l'horloge d'un de ces lieux retardait sur celle de l'autre et comme si aucune experience conceivable ne pouvait faire découvrir cette discordance".

[32] Stachel, John, "1905 and all that, How Einstein Claimed his Place in the Changing Landscape of Physics During his *Annus Mirablis*", 2005, Nature 433, pp. 215-217; p. 217.

[33] Poincaré, 1900c, pp. 458-459.

[34] In his 1898 paper, "The Measurement of Time", Poincaré propounded new ideas about time and distant simultaneity, Poincaré, 1898, p. 2: "We do not have the direct intuition of the equality of two intervals of time. People who believe they possess this intuition are dupes by an illusion". And then he continued to say, Poincaré, 1898, pp. 12-13: "We do not have the direct intuition of simultaneity, nor anymore of the equality of two durations. If we think we do have this intuition, it is but an illusion". However, Poincaré stated that we had replaced the non-intuition of simultaneity and the equality of two intervals of time with certain rules: time should be defined in such a way as to make the equations of classical mechanics, physics and astronomy as simple as possible.

[35] Poincaré, 1902/1968, p. 100. On page 92 of the 1904 German edition of *Science and Hypothesis*, *Wissenschaft und Hypothese*, there is a footnote redirecting the reader to endnote number 43 on pages 287-289. On page 92 Poincaré discusses time and distant simultaneity and the footnote refers to Poincaré's paper "la mesure du temps" and to the journal where one can find the paper: *Revue de métaphysique et de morale*. Endnote 43 at the end of the book supplies background technical material. *It does not discuss Poincaré's 1898 paper "Measurement of Time".*[35] Poincaré, Henri (1902/1904), *Wissenschaft und Hypothese*, Deutsch von F. und L. Lindemann Dritte Auflage, 1904, Leipzig und Berlin: B.G. Teubner, pp. 92; pp. 287-289.

The endnote first refers to the German 1889 edition of Newton's *Principia*, and then says that, Newton presupposed the existence of the "absolute time" [Newton ... setzte die Existenz einer "absoluten Zeit" voraus]. d'Alembert and Locke, for instance advanced the ancient relative time measurement [d'Alembert Locke u.a. hoben den relativen charakter alter zeitmaß hervor]. Then the reader is referred to an ancyclopidic article for historic information [vgl. die historischen Angaben bei A. Voss in dem Artikel uber die Prinzipien der rotation den Mechanik (Enzyklopädie der Math Wissenschaften. VI. 1).]. Subsequently one finds a mathematical explanation, "Perhaps the following analytical discussion may contribute to clarify […]" [Vielleicht kann die folgende analytische Erörterung hier zur klärung beitragen]. The physical-mathematical explanation that follows is not akin to any of Poincaré's explanations found in his papers. Poincaré, 1902/1904, pp. 92; 287-289. This endnote is similar to other complimentary endnotes for *Wissenschaft und hypohese*, most of which add technical explanations to Poincaré's simple explanations; the endnotes are quite unrelated to Poincaré's studies, and are incomprehensible to a lay reader. A few examples: In endnote 27 Clifford, Klein, Riemann, Lobatschewski, Hilbert, and Minkowski (*Geometrie der Zahlen* Heft, I Leipzig 1896) are mentioned. Poincaré, 1902/1904, p 276-279. In endnote 48 Newton, Helmholtz, Kirchhoff, and Thomson's writings are referred to. In addition there is reference to Pearson's *Grammar of Science* and Ernst Mach. Poincaré, 1902/1904, p 294. Endnote 74 discusses "Die Theorie der Elektronen" and supplies extensive references to relevant papers such as Lorentz's 1895 *Versuch*; and so do the following endnotes 75 and 76 supply a handful of sources. Poincaré, 1902/1904, pp. 316-317. Footnote 79 refers to Fizeau's experiment and references to Fizeau's 1859 paper, and to Michelson and Morley's repetition of the experiment (and reference to their paper). Poincaré, 1902/1904, p. 319. Note 93 discuss and refers to Maxwell's *Treatise on Electricity and Magnetism* of 1873. Poincaré, 1902/1904, p. 328. Maxwell's additional writings were mentioned in previous endnotes such as endnote 78. Poincaré, 1902/1904, pp. 318-319. And there is a "Nachtrag", which contains corrections and additional information and technical sources upon the endnotes that is burst upon the lay reader.

[36] Louis De Broglie distinguished between mathematical physics and theoretical physics, Broglie de, Louis, "Henri Poincaré et les théories de la physique", *Savants et Découvertes*, 1951, Paris: Albin Michel, p. 46: The first, according to De Broglie, is the profound and critical examination of the physical theories put forward by the researcher who assesses mathematical speculations in order to improve these theories and in order to render their inherent proofs more rigorous. In contrast, theoretical physics is the construction of theories suitable to serve as an explanation of the experimental facts and to guide the work of the laboratory staff. Extensive mathematical knowledge is a pre-requisite, although it is not, ordinarily, the work of real mathematicians; it requires wide knowledge of the experimental facts, and mainly some kind of intuition in physics, which not all mathematicians have, as did Poincaré. Poincaré, according to De Broglie, was especially destined to engage fruitfully in mathematical physics. De Broglie asserted that Einstein was a scientist of the second type, and therefore he had succeeded in arriving at the theory of relativity. These definitions stem from the French scientific tradition, and, in respect to Poincaré, was even apparent in the manner in which Poincaré presented his scientific endeavors.

[37] Poincaré, 1905c, p. 4.

[38] Poincaré, 1905c, p. 4.

[39] Poincaré, 1905c, p. 5.

[40] Poincaré, 1905c, p. 5.

[41] Poincaré, 1905c, p. 5.

[42] Poincaré, 1905c, p. 5.

[43] Poincaré, 1905c, p. 5.

[44] Poincaré, 1905c, p. 7.

[45] Poincaré, 1905c, p. 18.

[46] Poincaré, 1905c, pp. 16-17.

[47] Poincaré, 1905c, p. 18.

[48] Pierseaux, 2004, p. 62.

[49] Poincaré to Lorentz, May, 1905, Poincaré Archives Nancy; Letter Number 38.4., Walter, Bolmont, and Coret (ed), 2000, p. 257.

[50] Poincaré, 1905c, p. 18.